\documentclass[12pt,a4paper,final]{iopart}

\usepackage{iopams}  
\usepackage{graphicx}
\usepackage[breaklinks=true,colorlinks=true,linkcolor=blue,urlcolor=blue,citecolor=blue]{hyperref}
\usepackage{epstopdf}
\begin{document}

\title[Temperature and composition dependence of short-range order, entropy and bonds]{Temperature and composition dependence of short-range order and entropy, and statistics of bond length: the semiconductor alloy (GaN)$_{1-x}$(ZnO)$_x$}

\author{Jian Liu$^{1}$, Luana S. Pedroza$^{1}$, Carissa Misch$^{2}$, Maria V. Fern\'{a}ndez-Serra$^{1}$, Philip B. Allen$^{1}$}
\address{$^1$Department of Physics and Astronomy, Stony Brook University, Stony Brook, NY 11794-3800, United States.}
\address{$^2$Smith College, Northampton, MA 01063, United States.}
\eads{\mailto{Jian.Liu@stonybrook.edu}, \mailto{philip.allen@stonybrook.edu}}

\begin{abstract}
We present total energy and force calculations on the (GaN)$_{1-x}$(ZnO)$_{x}$ alloy. Site-occupancy configurations are generated by Monte Carlo (MC) simulations, based on a cluster expansion (CE) model proposed in a previous study.  Surprisingly large local atomic coordinate relaxations are found by density-functional calculations using a 432-atom periodic supercell, for three representative configurations at $x=0.5$.  These are used to generate bond length distributions.  The configurationally averaged composition- and temperature-dependent short-range order (SRO) parameters of the alloys are discussed. Entropy is approximated in terms of pair distribution statistics and thus related to SRO parameters. This approximate entropy is compared with accurate numerical values from MC. An empirical model for the dependence of bond length on local chemical environments is proposed.
\end{abstract}

\vspace{2pc}
\submitto{\JPCM}

\section{Introduction}

The pseudobinary (GaN)$_{1-x}$(ZnO)$_{x}$ semiconductor alloy is promising as a photocatalyst using visible light for splitting water into O$_2$ and protons\cite{Maeda1}\cite{Maeda2}.  A co-catalyst ({\it e.g.} an oxide of Rh or Cr) can then reduce the protons to create H$_2$ fuel. High efficiency is partly attributed to the fact that the band gap and other physical properties can be tuned by varying the composition of the alloy. Compositional fluctuations lead to surprisingly strong dependence of bond length on local chemical environment.  Localization of states induced by disorder makes carrier transport in the alloy significantly different from the constituent semiconductors\cite{ZungerCarrier}\cite{Wei}.  Understanding the role of order and disorder in determining physical properties such as bond length, is complicated by the difficulty in modelling the numerous local chemical environments of the alloy.  Theoretical approaches assuming effective medium (coherent potential approximation, CPA)\cite{Gyorffy1}\cite{Gyorffy2}\cite{CPA}, periodic virtual crystal with the actual configuration treated as perturbations (virtual crystal approximation, VCA), and random alloy with special quasirandom structure (SQS)\cite{SQS} mimicking the correlation function, may not be sufficient, especially because of large deviations from complete disorder.  Increased computational power now allows accurate computational prediction of short-range order (SRO) using density-functional theory (DFT).   Thus long-standing problems of concern to B. L. Gyorffy\cite{Gyorffy3}\cite{Gyorffy4} can start to be addressed.  In our previous study\cite{LL}, refered to hereafter as \uppercase\expandafter{\romannumeral1}, strong SRO is predicted in the (GaN)$_{1-x}$(ZnO)$_{x}$ alloy.  The present study addresses various physical properties.  SRO is taken fully into account, by averaging within realistic thermodynamic ensembles.\\
\newline
For the effect of SRO, extensive theoretical studies can be found on pseudobinary $(A_{1-x}B_x)C$ semiconductor alloys\cite{Thorpe2}\cite{Zunger1}\cite{Zunger2}\cite{ternary}\cite{quasiparticle} where only the cation sites are substitutional and $A, B$ are isovalent elements. Quaternary isovalent \uppercase\expandafter{\romannumeral3}-\uppercase\expandafter{\romannumeral5} semiconductor alloys $A_{1-x}B_xC_{1-y}D_{y}$ are also studied by assuming the additivity of atomic natural radii\cite{Thorpe1}. SRO is assumed either explicitly or implicitly to be weak in previous studies. The nature of the (GaN)$_{1-x}$(ZnO)$_{x}$ semiconductor alloy, however, is complicated mainly by two facts, namely the cation and the anion sublattices are interpenetrating, and the constituent GaN and ZnO semiconductors differ in valence. The present study has three aims. (1) To extend the study of SRO in \uppercase\expandafter{\romannumeral1}. (2) To compute the configurational entropy of disorder and relate it to the Warren-Cowley measure of SRO. (3) To observe and describe the relatively large bond-length fluctuation associated with variation in local chemical environments and biased by SRO.\\

\section{Method}

In \uppercase\expandafter{\romannumeral1} the (GaN)$_{1-x}$(ZnO)$_{x}$ semiconductor alloy was modeled in the wurtzite structure with cation and anion sublattices.  Only Ga or Zn atoms occupy the cation sublattice, and N or O atoms occupy the anion sublattice. A cluster expansion (CE)\cite{CE1}\cite{CE2}\cite{CE3} model was constructed.  Ising parameters $\sigma_i$  describe the $i^{\rm th}$ site occupancy, with $\sigma=1$ denoting Ga and N and $\sigma=-1$ denoting Zn and O. The formation energy of the alloy was then expanded in terms of zero-, one- and two-body Ising-type interactions.  Three-body Ising interactions were included in the original fit, and found to be unimportant.  The effective cluster interactions were found by fitting to a data base of random alloy super-structures, with unit cells of various sizes, up to 196 atoms.  The formation energy was computed using well-converged DFT, including full relaxation of local atomic coordinates.  These calculations used the VASP codes\cite{Kresse} with the PBE functional\cite{PBE}, and included Ga and Zn 3d states as valence states.  Empirical on-site U parameters of 3.7 eV and 6.0 eV were used for the Ga and Zn 3d states.  The $ATAT$ package\cite{ATAT1}\cite{ATAT2}\cite{ATAT3} was used to adjust cluster parameters to fit the formation energy.  The process was iterated, by using the first set of cluster parameters, $via$ MC, to generate new structures, no longer random, but characteristic for temperatures of $\approx$1500K.  New DFT energies and lattice relaxations gave an enlarged and improved data base, which then generated a final set of improved CE parameters.\\
\newline
The $ATAT$ code was then used, in I, to compute by MC calculations the theoretical $(x,T)$ phase diagram of the alloy.  In a narrow composition range near $x=1/2$, an ordered GaNZnO phase is predicted to be thermodynamically stable below $T_c=870$K.  Long-range order disappears in a strongly first-order fashion as $T$ increases past $T_c$.  Actual alloys are made at higher $T$.  Slow cooling through $T_c$ does not generate the well-ordered low $T$ phase, because ionic diffusion is too slow at this temperature to permit equilibration.\\
\newline
Now we exploit the CE to further study thermodynamic properties, and also the structural properties of larger unit cells.  The ensembles are equilibrated at particular choices of $(x,T)$. We choose a $6\times6\times3$ supercell (432 atoms) with periodic boundary conditions, as a compromise between computational expense and statistical realism. Such a supercell can exhibit a wide range of local environments. For each $(x,T)$, 200,000 MC passes were used for equilbration.  Then 100 configurations were chosen, each separated from the previous configuration by 1,000 passes.  These are used for further SRO-related analysis. Monte Carlo (MC) simulations only generate site occupancy, and do not predict the actual lattice relaxations.  Subsequent DFT total energy and force calculations are required to relax the atomic coordinates. Three representative configurations equilibrated at $1200$K for the $x$=0.5 composition are chosen for structural relaxation, which is computationally expensive on 432-atom lattices.  The structural relaxations are done using DFT with the generalized-gradient approximation (GGA). The Perdew-Burke-Ernzerhof (PBE)\cite{PBE} version of the exchange-correlation functional is used. Kohn-Sham wavefunctions are expanded in the basis of linear combinations of localized atomic orbitals, as implemented in the {\sc SIESTA} code\cite{SIESTA}. We use a variationally optimized double-$\zeta$ polarized (DZP) basis set. Ga- and Zn- 3d electrons are treated explicitly as valence electrons.  No empirical ``+U'' was used for Ga or Zn d-electrons.  The cell volume is also relaxed.\\ 
\newline
In figure 1, we show the energy $(E)$ distribution $P(E)$ of the $432$-atom thermodynamic ensemble equilibrated at temperature $T=1200$K. The distribution at a hypothetical temperature $T=10,000$K is also shown for comparison, because it represents an approximately random alloy.   In both cases, the width of the distribution is approximately $\bar{E}/\sqrt(432)$.  The mean energy $\bar{E}$ is well below $k_B T$, because the spectrum of cluster-expansion energies is bounded above, with median value not far above 0.1eV per atom.  Larger cell sizes would narrow $P(E)$, but should not affect the mean $\bar{E}$ since the rare long-range ordered clusters (some forbidden and some encouraged by 432-atom periodicity) are very unlikely to skew the energy distribution.
\begin{figure}[htb!]
 \centering
 \includegraphics[scale=0.15]{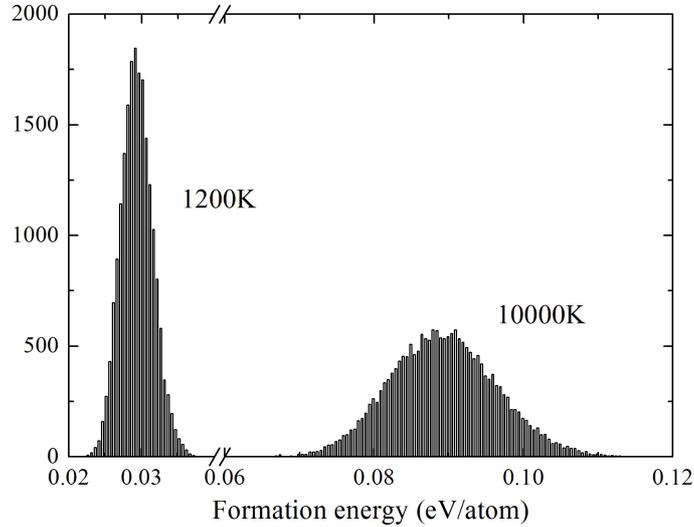}
 \caption{Energy distribution $P(E)$ for the 432-atom supercell at $x$=0.5, equilibrated at $T=1200$K and $T=10,000$K.  Three representative 1200K samples at $\bar{E}=0.0295$ and $\bar{E}\pm 0.0017$ eV/atom were further relaxed and used for the present study.  The 10,000K distribution  represents an approximately random alloy.}
\end{figure}\\
\begin{figure}[htb!]
 \centering
 \includegraphics[scale=0.18]{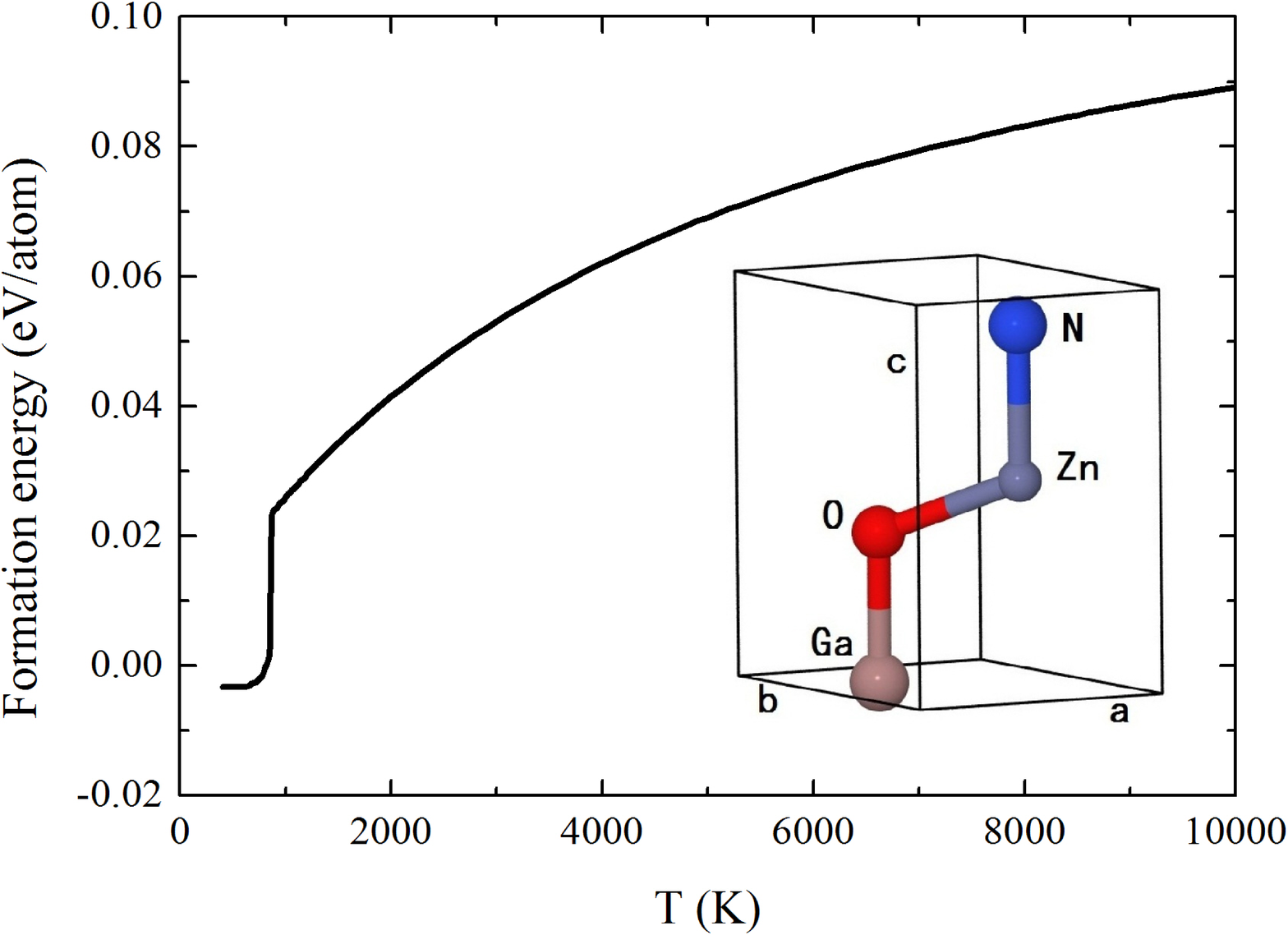}
 \caption{Formation energy as a function of temperature for $x$=0.5.  Statistical errors, smaller than the width of the line, are discussed in the text.  The latent heat for the order-disorder transition is about 0.026 eV/atom. At 1200K, the contribution to specific heat from thermal compositional disordering is about 0.2$k_B$ per atom. The inset shows the pseudo-wurtzite unit cell, with atoms ordered as in the $T=0$ prediction (ref. \cite{LL}) from MC simulations.}
\end{figure}
\newline
Figure 2 shows the thermally averaged formation energy as a function of temperature.  The thermal ensembles of configurations are similar to those of Fig. 1.   The statistical error from averaging over 100 statistically-independent MC samples is less than the thickness of the line.  The systematic error from the 432-atom periodicity is probably very small.  As found in I, at fifty-fifty mixing, the alloy undergoes a first-order phase transition at $T\approx 870$K. The ordered structure at $x$=0.5 corresponds to the periodic replication of the unit cell shown in the inset. The formation energy for the ordered structure is about $-3$ meV/atom, barely stabilized against phase separation. The CE parameters\cite{LL} show that first-neighbor cation-anion pairs strongly prefer to retain the proper valence pairing (Ga to N and Zn to O), but second neighbor interactions prefer to switch valences (Zn to Ga and O to N).  The ordered phase at $x=0.5$ is a compromise.   In the wurtzite lattice, each cation or anion site is surrounded by 12 second-neighbor cation or anion sites.  Six of these (referred to as type-$ab$) are in the same hexagonal $ab$ plane  and 6 others (referred to as type-$c$) are in different $ab$ planes above and below the reference site.  The 6 type-$ab$ second neighbors are unable to satisfy valence switching, because of the stronger need for proper first-neighbor valences.  But the 6 type-$c$ second neighbors are all able to satisfy valence switching, at the expense of forcing one out of four first neighbors to have improper valence (Ga-O or Zn-N rather than the preferred Ga-N and Zn-O).  Three representative configurations (at $\bar{E}=0.0295$ and $\bar{E}\pm 0.0017$ eV/atom) are chosen from the $T=1200$K ensemble for further structural relaxations.\\

\section{Results and discussions}

\subsection{Short-Range Order and Entropy}

Competition between energetic preference and entropic dislike for particular local chemical environments, is what determines the alloy SRO.  In (GaN)$_{1-x}$(ZnO)$_{x}$ alloy, valence-matched pairs (Ga-N and Zn-O) are energetically favored over unlike-pairs (Ga-O and Zn-N), since charge is better balanced in like-pairs. To quantify the degree of SRO in the alloy, we generalize the commonly used Warren-Cowley SRO parameters\cite{Cowley1}\cite{Cowley2}.  For a simple binary $A_{1-x}B_{x}$ alloy, these are defined as
\begin{equation}
\gamma_i=1-\frac{P_i(B \vert A)}{x}
\end{equation}
where $P_i(B \vert A)$ is the conditional probability of finding a B atom in the $i$th shell centered around an A atom. Values $\gamma_i<0$ indicate preference for unlike-pairs, while $\gamma_i>0$ indicate preference for like-pairs. Suppose SRO parameters $\gamma_{i=1,2,...,n}$ are known.  Then the configurational entropy for the simple binary alloy\cite{Onabe} can be approximated as
\begin{eqnarray}
\fl S=S_{\rm rand}-\{S_{\rm rand}\sum_{i=1}^nq_i+k_B\sum_{i=1}^n\frac{q_i}{2}[x(x+y\gamma_i)\ln x(x+y\gamma_i)+\nonumber\\
y(y+x\gamma_i)\ln y(y+x\gamma_i)+2xy(1-\gamma_i)\ln xy(1-\gamma_i)]\}
\end{eqnarray}
where $y=1-x$, $S_{\rm rand}=-k_B(x\ln x + y\ln y)$ is the configurational entropy for a random alloy, and $q_i$ is the coordination number of the $i$th shell. For example, $q_1$ corresponds to the four-fold coordinated first-neighbor shell, and $q_2$ corresponds to the twelve-fold coordinated second-neighbor shell.  The term in the bracket gives the loss of entropy due to deviation from the random structure caused by SRO.  A derivation of this formula is given in the Appendix.\\
\newline
Eqn. (2) can be generalized to the (GaN)$_{1-x}$(ZnO)$_{x}$ semiconductor alloy, by proper generalization of the SRO parameters.    Cation and anion sublattices are independent and interpenetrating, each complying with the occupancy rule, namely cation sites are occupied by Ga or Zn and anion sites are occupied by N or O.  There are three types of pair-wise Ising interactions, (i) cation-cation, (ii) anion-anion, and (iii) cation-anion.  The SRO parameters are defined separately.   Cation-cation parameters are defined as $\beta_i=1-\frac{P_i(\rm Zn \vert Ga)}{m_{\rm Zn}}$, and anion-anion parameters as $\gamma_i=1-\frac{P_i(\rm O \vert N)}{m_{\rm O}}$.  These are just Warren-Cowley parameters; $m_A$ is the concentration ($x$ or $1-x$) of species $A$.  Although the second-neighbor type-$ab$ and type-$c$ distances are negligibly different, SRO parameters must be defined separately for type-$ab$ and type-$c$.  For example, in the ordered ground state, if no distinction is made between type-$ab$ and type-$c$, the SRO parameters $\beta_2$  and $\gamma_2$ both equal 0, whereas $\beta_2=\gamma_2=0$ should designate random occupancy. Therefore, the SRO parameters are defined separately for type-$ab$ and type-$c$.  The subscripts $ab$ and $c$ indicate type-$ab$ and type-$c$ pairs. For example, $q_{2,ab}=6$ and $q_{2,c}=6$.   In the $x=0.5$, $T=0$ ordered ground state, $\beta_{2,ab}=\gamma_{2,ab}=1$, whereas $\beta_{2,c}=\gamma_{2,c}=-1$.\\

For (iii), cation-anion pairs differ in that cations can not occupy the anion sublattice and vice versa.  Only one SRO parameter $\alpha_i$ is needed for enumerating the cation-anion pair probability distribution for the $i^{\rm th}$ shell, since we have the relations
\numparts
\begin{eqnarray}
N_{i,\rm GaN}+N_{i,\rm GaO}&=q_iNx\\
N_{i,\rm ZnN}+N_{i,\rm ZnO}&=q_iN(1-x)\\
N_{i,\rm GaO}+N_{i,\rm ZnO}&=q_iN(1-x)\\
N_{i,\rm GaN}+N_{i,\rm ZnN}&=q_iNx,
\end{eqnarray}
\endnumparts
where $N_{i,{\rm AB}}$ denotes the number of AB pairs in the $i^{\rm th}$ cation-anion shell.  The relations $N_{i,\rm GaN}-N_{i,\rm ZnO}=q_iN(2x-1)$ and $N_{i,\rm GaO}=N_{i,\rm ZnN}$ follow directly. Positive $\alpha_i$ indicates preference for GaN and ZnO pairs, while negative $\alpha_i$ indicates preference for GaO and ZnN pairs.  The pair probabilities $p_{i,\rm AB}$ are related to the SRO parameters $\alpha_i$ by
\numparts
\begin{eqnarray}
p_{i,\rm GaO}=m_{\rm Ga}(m_{\rm O}-m_{\rm O}\alpha_i)\\
p_{i,\rm GaN}=m_{\rm Ga}(m_{\rm N}+m_{\rm O}\alpha_i)\\
p_{i,\rm ZnO}=m_{\rm Zn}(m_{\rm O}+m_{\rm N}\alpha_i)\\
p_{i,\rm ZnN}=m_{\rm Zn}(m_{\rm N}-m_{\rm N}\alpha_i),
\end{eqnarray}
\endnumparts
Similar to cation-cation (or anion-anion) pairs, for the four nearest neighbors in the first shell, the cation-anion pair along the $c$-axis is distinguished from the other three. Now $q_{1,ab}=3$ and $q_{1,c}=1$. Finally, the SRO-corrected entropy, by generalizing the derivation in the appendix, is approximated by
\begin{eqnarray}
\fl S=S_{\rm rand}-\sum_i S_{\rm rand} (q_{1,i}+q_{2,i})-k_B\sum_i\{\nonumber\\
\frac{q_{2,i}}{4}[2m_{\rm Ga}m_{\rm Zn}(1-\beta_i)\ln m_{\rm Ga}m_{\rm Zn}(1-\beta_i)+\nonumber\\
m_{\rm Ga}(m_{\rm Ga}+m_{\rm Zn}\beta_i)\ln m_{\rm Ga}(m_{\rm Ga}+m_{\rm Zn}\beta_i)+\nonumber\\
m_{\rm Zn}(m_{\rm Zn}+m_{\rm Ga}\beta_i)\ln m_{\rm Zn}(m_{\rm Zn}+m_{\rm Ga}\beta_i)+\nonumber\\
2m_{\rm N}m_{\rm O}(1-\gamma_i)\ln m_{\rm N}m_{\rm O}(1-\gamma_i)+\nonumber\\
m_{\rm N}(m_{\rm N}+m_{\rm O}\gamma_i)\ln m_{\rm N}(m_{\rm N}+m_{\rm O}\gamma_i)+\nonumber\\
m_{\rm O}(m_{\rm O}+m_{\rm N}\gamma_i)\ln m_{\rm O}(m_{\rm O}+m_{\rm N}\gamma_i)]+\nonumber\\
\frac{q_{1,i}}{2}[m_{\rm Ga}(m_{\rm N}+m_{\rm O}\alpha_i)\ln{m_{\rm Ga}}(m_{\rm N}+m_{\rm O}\alpha_i)+\nonumber\\
m_{\rm Ga}(m_{\rm O}-m_{\rm O}\alpha_i)\ln{m_{\rm Ga}}(m_{\rm O}-m_{\rm O}\alpha_i)+\nonumber\\
m_{\rm Zn}(m_{\rm O}+m_{\rm N}\alpha_i)\ln{m_{\rm Zn}}(m_{\rm O}+m_{\rm N}\alpha_i)+\nonumber\\
m_{\rm Zn}(m_{\rm N}-m_{\rm N}\alpha_i)\ln{m_{\rm Zn}}(m_{\rm N}-m_{\rm N}\alpha_i)]\},
\end{eqnarray}
where $i$ runs over neighbor shells. This approximate entropy is completely determined by the SRO parameters. It can be tested against the theoretical configurational entropy obtained during the MC simulations. This is extracted for constant volume from the $\bar{E}(T)$ curve (figure 2) using $S=\int{d\bar{E}/T}$. In figure 3, we show the temperature-dependent SRO parameters for the $x$=0.5 case. Each SRO parameter is obtained by averaging over 100 configurations equilibrated at the corresponding temperature in MC simulations. As expected, the SRO parameters decrease with increasing temperature.  The nearest neighbor (NN) SRO parameters, both first (1NN) and second (2NN) are positive, indicating the energetic preference for Ga-N, Zn-O, Ga-Ga and Zn-Zn pairs. The alloy is more short-range ordered in the hexagonal plane than in the $c$-axis direction. The 1NN SRO parameters are much larger than the 2NN SRO parameters, and decay more slowly at elevated temperatures. At temperatures below $2000$K, the energetic preference for Ga-N and Zn-O pairs is remarkable, which further verifies the necessity of treating SRO explicitly in the modelling. Using the obtained SRO parameters, we are then able to calculate the SRO-corrected configurational entropy. The result is shown in figure 4. The SRO-corrected configurational entropy is systematically larger than the true value obtained from MC simulations, due to the inter-dependence of SRO parameters. The error becomes smaller with increasing temperature.\\
\begin{figure}[htb!]
 \centering
 \includegraphics[scale=0.20]{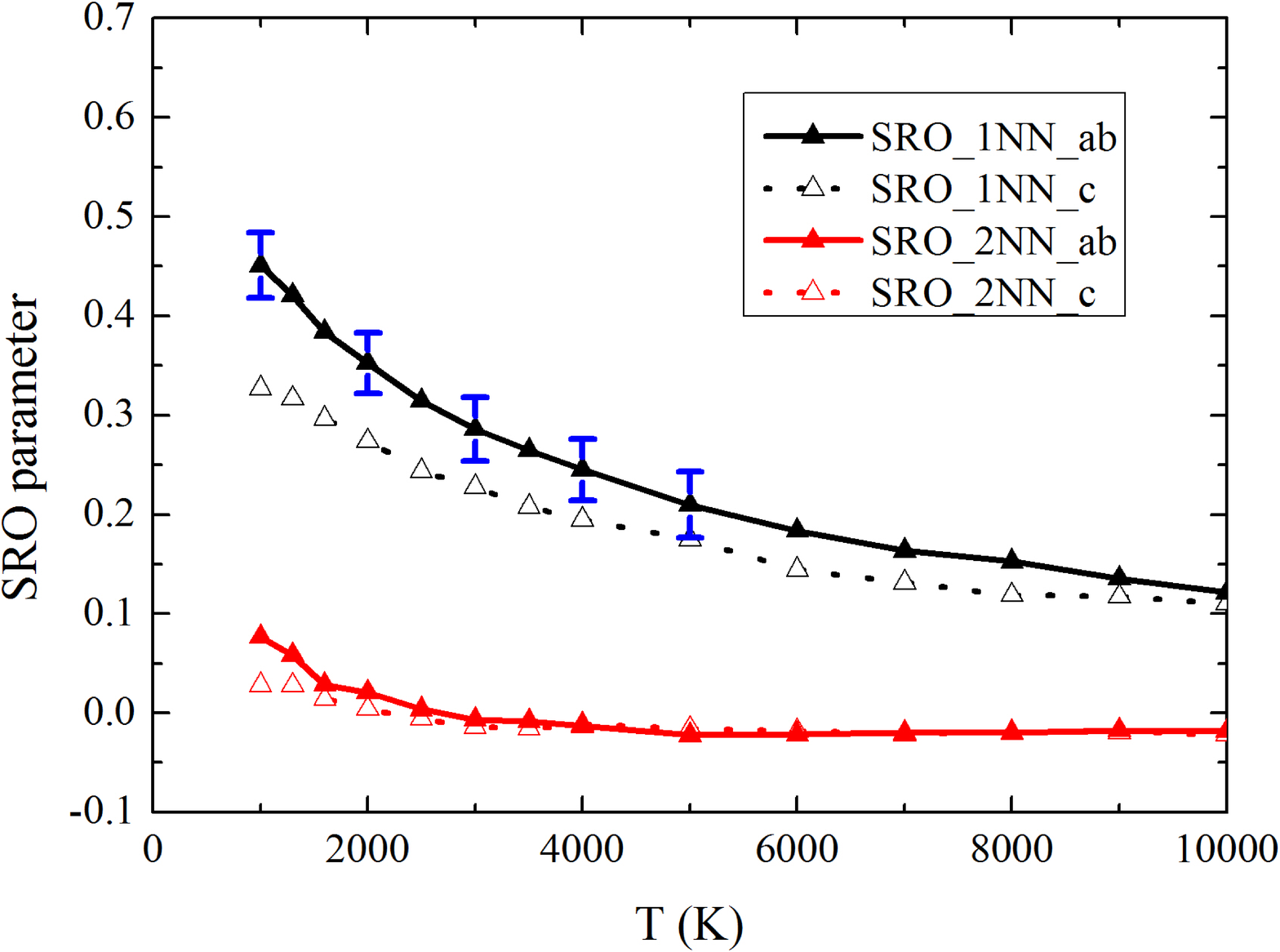}
 \caption{Temperature dependence of short-range order parameters from MC simulations at $x$=0.5. Only 1NN and 2NN SRO parameters are shown.  Error bars represent the rms width of the distribution.  The statistical error from averages over 100 statistically-independent samples drawn from this distribution is $\sqrt(100)$ times smaller.}
\end{figure}
\begin{figure}[htb!]
 \centering
 \includegraphics[scale=0.20]{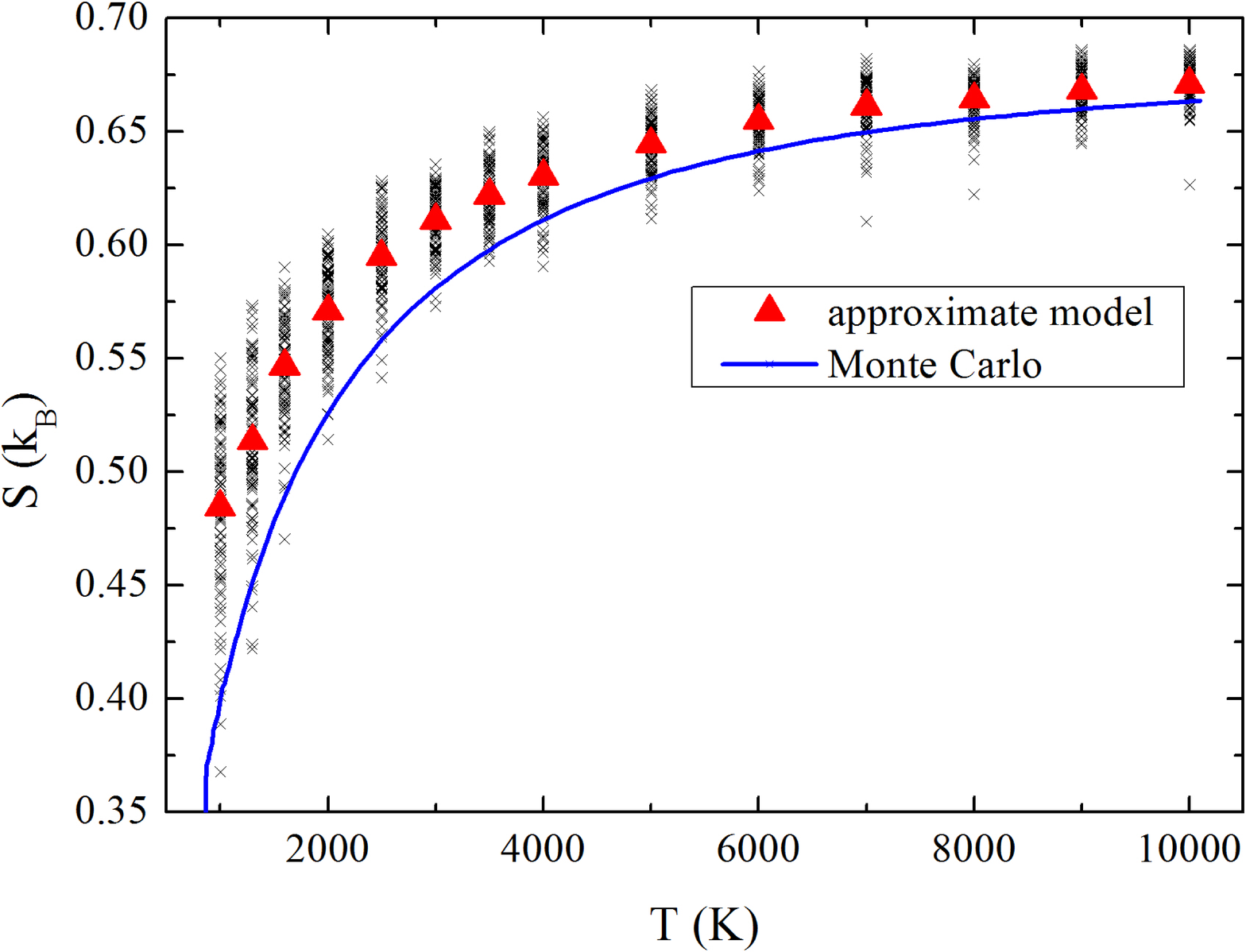}
 \caption{SRO-corrected approximate configurational entropy at $x$=0.5, calculated according to formula (5).   Black crosses are values for the 100 particular configurations of the statistical distribution.  Red triangles are the averaged entropy, with statistical error $\approx 1/\sqrt(100)$ of the width of the distribution of crosses. As the alloy becomes increasingly random at high $T$, the configurational entropy approaches the limit $k_B\ln2$.  The solid curve gives true MC values.}
\end{figure}\\
The composition-dependent 1NN SRO parameters and configurational entropy are shown in figure 5 and figure 6 respectively. Upon mixing, the configurational entropy deviates further from that of the random alloy. The maximum of configurational entropy occurs in the equal-mixing composition. For SRO parameters, similar behavior is observed above $1500$K. Below $1500$K, the maxima are biased away from the equal-mixing in both higher and lower composition directions. This bimodal character is also observed in a study of ordering of ternary nitride semiconducting alloys\cite{ternary}.\\
\begin{figure}[htb!]
 \centering
 \includegraphics[scale=0.20]{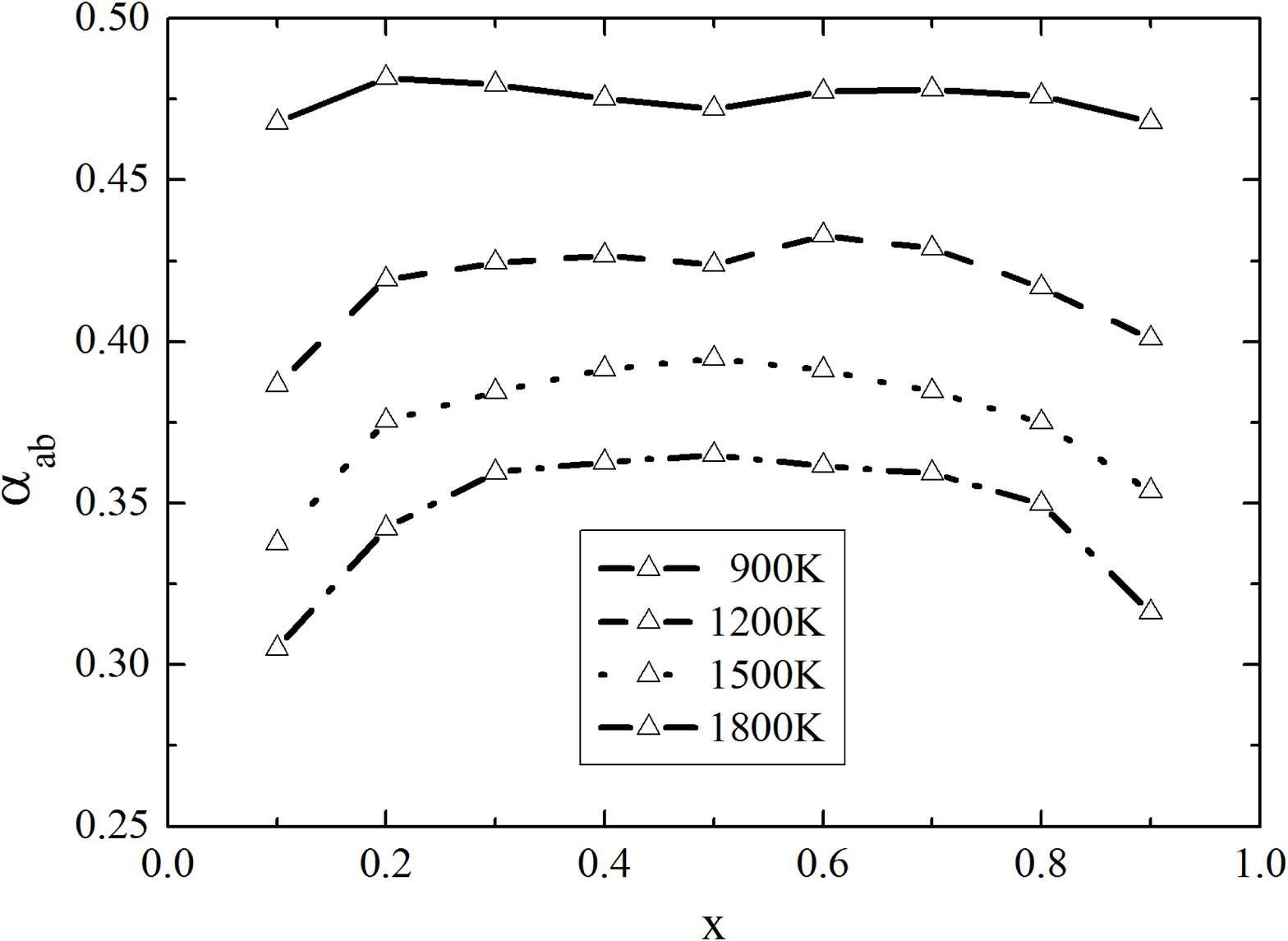}
 \caption{Composition- and temperature-dependent 1NN SRO parameters.  Statistical errors are small, similar to Fig. 3.  2NN SRO parameters are not shown for clarity since they are rather small and subject to more significant statistical errors.}
\end{figure}
\begin{figure}[htb!]
 \centering
 \includegraphics[scale=0.20]{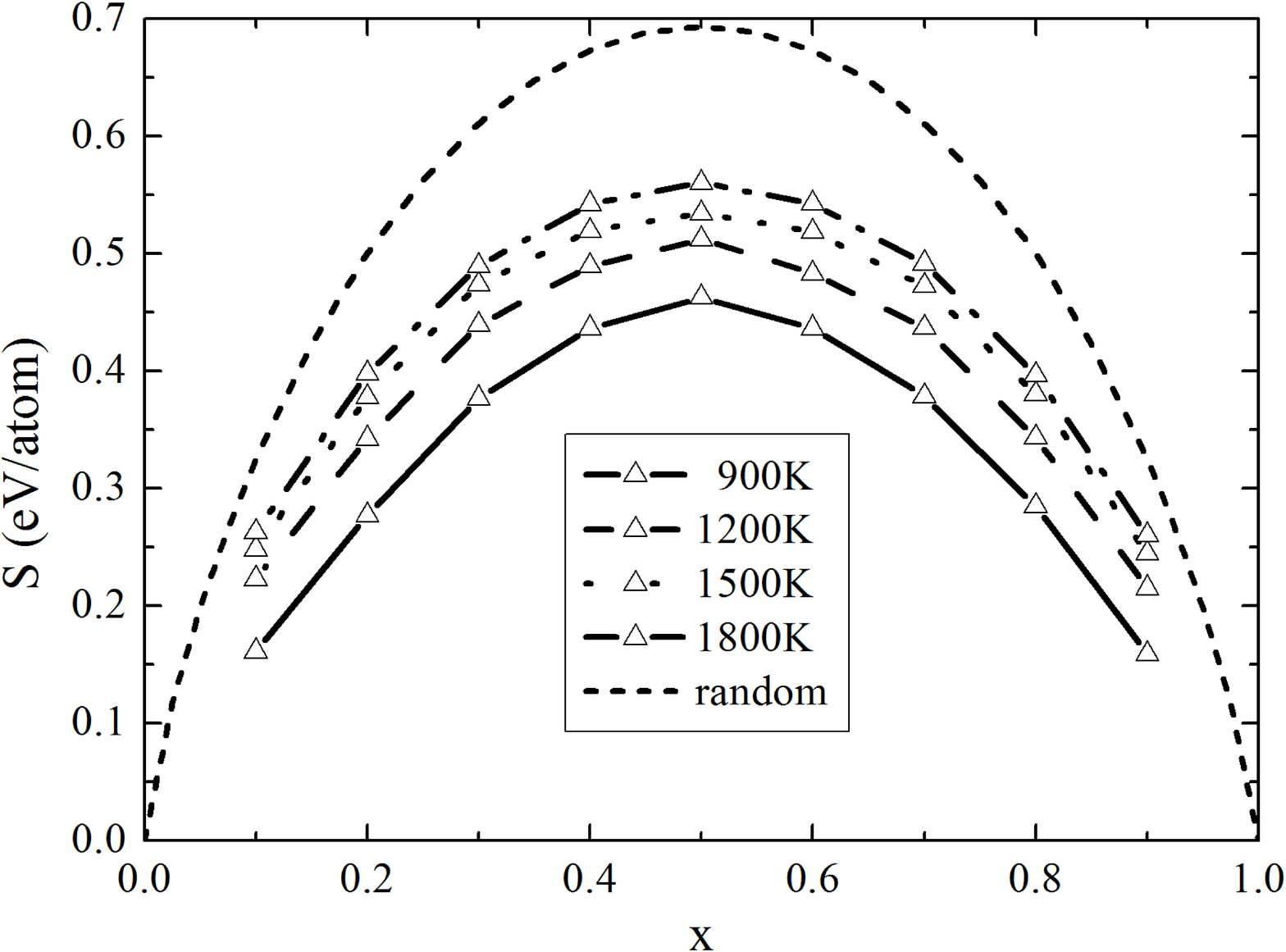}
 \caption{Composition- and temperature-dependent configurational entropy (from MC, not from the approximate model.)  Statistical errors are small, similar to those of Fig. 4.  The dashed line marks the configurational entropy for a random alloy.}
\end{figure}

\subsection{Statistics of bonds and bond length}

The SRO is dominated by non-random NN cation-anion pairs.  Now let us look in more detail at the local environments of the NN pairs. We define the notation $P$(AB)$_{mn}$ to denote the probability of a NN AB pair in which the A atom has $1\le m\le 4$ neighboring B atoms and the B atom has $1\le n\le 4$ neighboring A atoms. Figure 7 shows the temperature-dependence of the GaN pair probability $P(\rm GaN)_{mn}$ at $x$=0.5, when the Ga of the pair has a total of $m$ N first neighbors and the N of the pair has a total of $n$ Ga first neighbors. The pair probability resembles a two-dimensional binomial-like distribution. The SRO parameters reduce with increasing temperature, resulting in smoother pair distributions. Figure 8 shows the composition-dependence of the GaN pair probability at $T=1800$K. Upon mixing, the peak shifts in the unlike-pair direction. Based on these observations, we approximate the pair distribution with the binomial function (using GaN as an example):
\begin{eqnarray}
\fl P(\rm GaN)_{mn}=m_{\rm Ga}C_4^m(m_{\rm N}+m_{\rm O}\alpha)^m(m_{\rm O}-m_{\rm O}\alpha)^{4-m}\times\nonumber\\
C_4^n(m_{\rm Ga}+m_{\rm Zn}\alpha)^n(m_{\rm Zn}-m_{\rm Zn}\alpha)^{4-n}
\end{eqnarray}
where $C_4^m$ denotes the binomial coefficient ${4!}/{m!(4-m)!}$. In figure 9, we compare the GaN pair probability obtained in MC with the proposed binomial distribution. The binomial distribution approximates the real distribution qualitatively. However, due to the interpenetrating character between clusters in the sublattice, the applicability of the binomial distribution becomes questionable upon mixing.\\ 
\begin{figure}[htb!]
 \centering
 \includegraphics[scale=0.35]{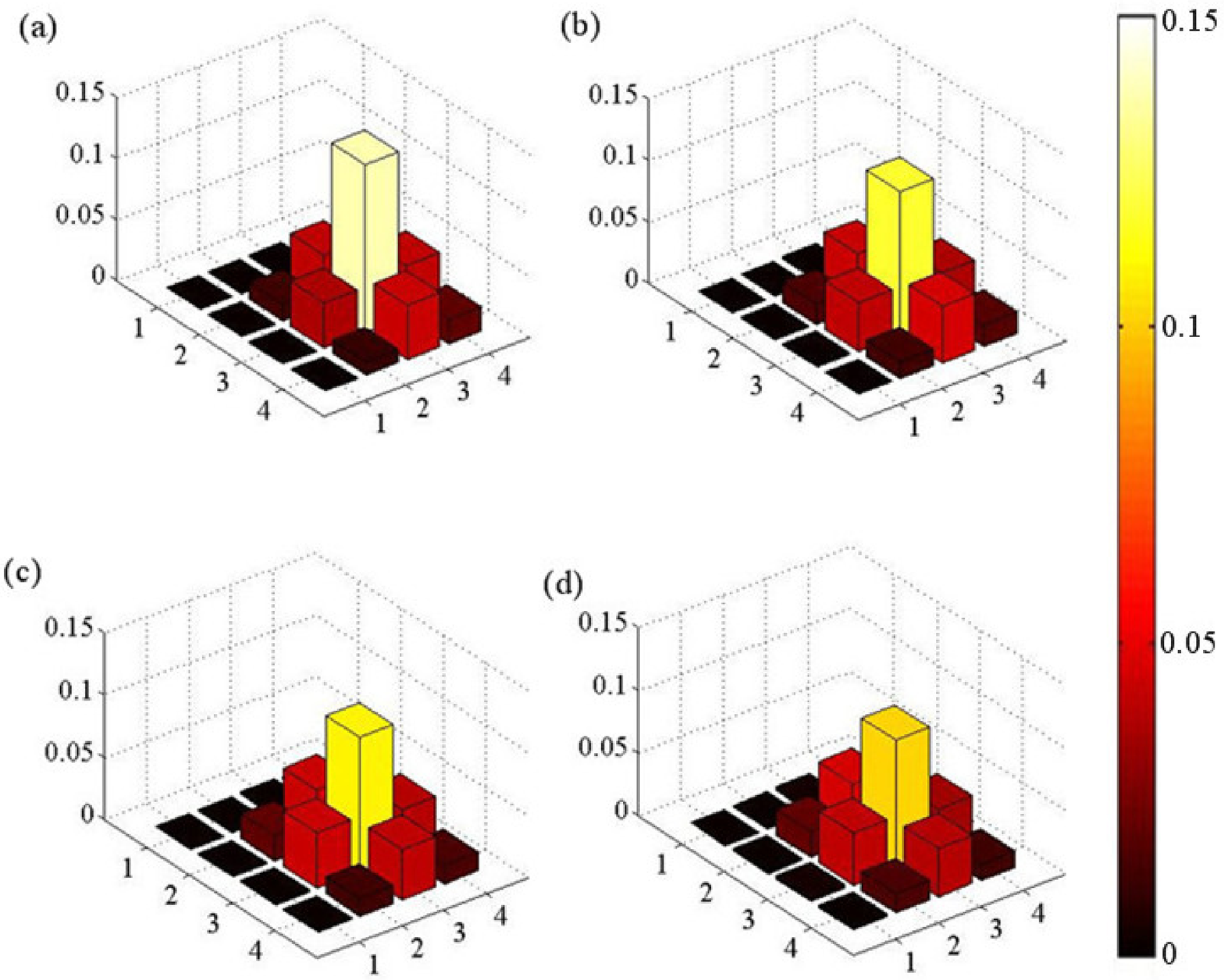}
 \caption{The GaN pair probability $P(\rm GaN)_{mn}$ for $x$=0.5 at temperature (a) $900$K, (b) $1200$K, (c) $1500$K, (d) $1800$K.}
\end{figure}
\begin{figure}[htb!]
 \centering
 \includegraphics[scale=0.35]{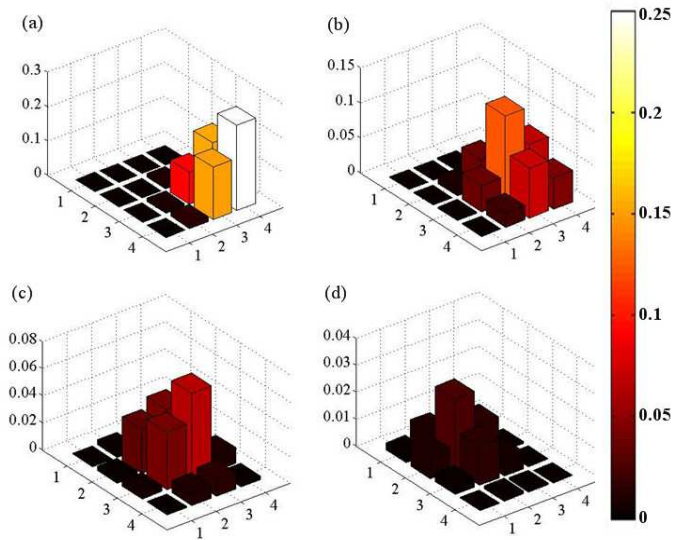}
 \caption{The GaN pair probability $P(\rm GaN)_{mn}$ at $T=1800$K for composition $x$ (a) 0.2, (b) 0.4, (c) 0.6, (d) 0.8.}
\end{figure}
\begin{figure}[htb!]
 \centering
 \includegraphics[scale=0.35]{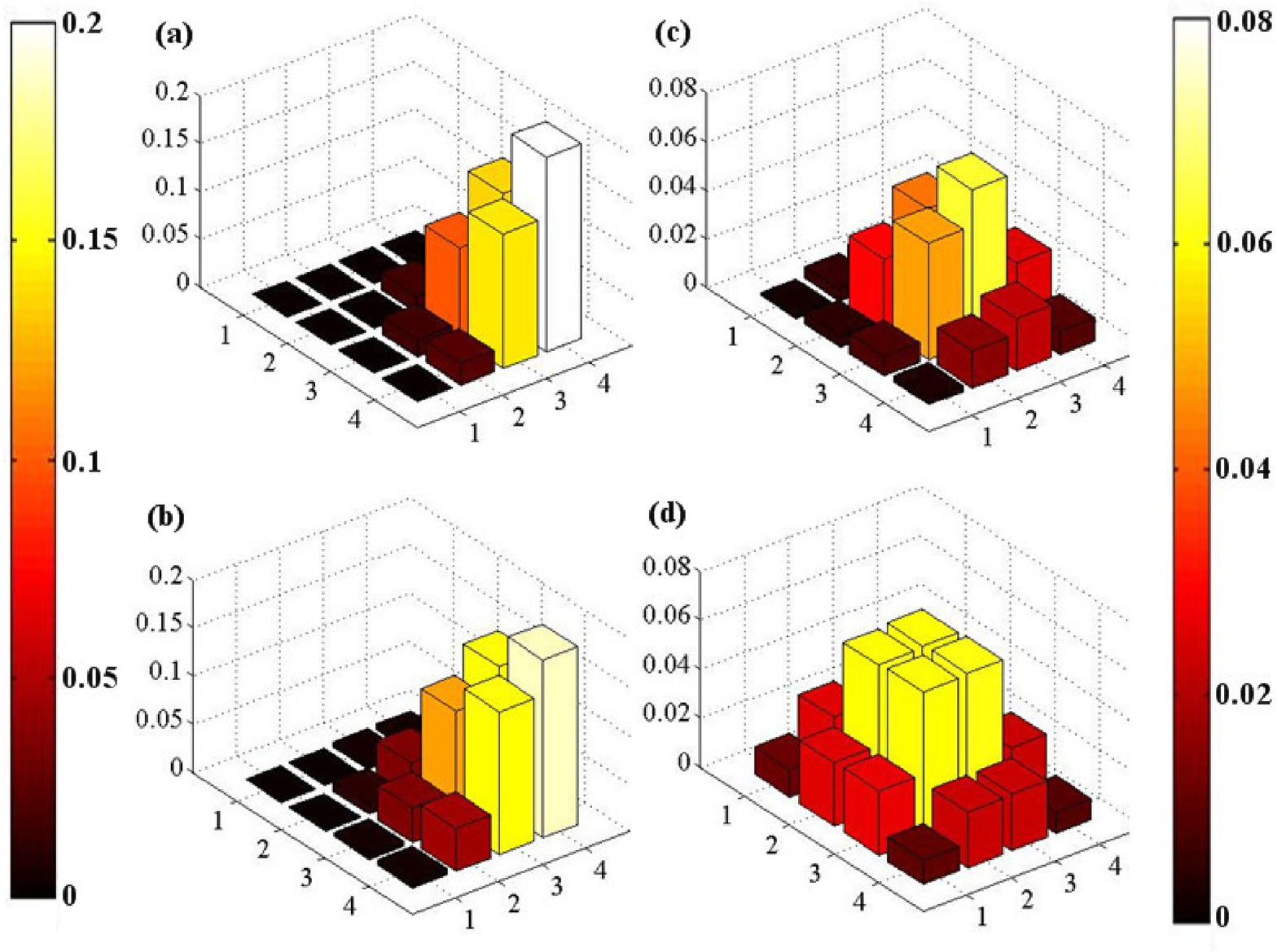}
 \caption{Comparison of $P(\rm GaN)_{mn}$ from binomial distribution with MC simulations at a hypothetical temperature $T=5000$K for $x$=0.2 (a, b) and $x$=0.5 (c, d). The upper figures (a, c) are from MC simulations, and the lower figures (b, d) are from Eq. 6.}
\end{figure}\\
For the statistics of bond length, and in particular their dependence on local chemical environment, we use the three representative configurations for $x=0.5$ and $T=1200$K, computing relaxed atomic coordinates using DFT.  In isovalent semiconductor alloys, additivity of atomic radii is a good assumption\cite{Thorpe1}. However, additivity is not applicable for the (GaN)(ZnO) alloy. In figure 10, we show results of structural relaxations. A surprising amount of bond length heterogeneity persists even when particular first neighbor environments $(mn)$ are treated separately. It is convenient to redefine the indices $(mn)$, now called $(pq)$, of bond lengths $d({\rm AB})_{pq}$ so that $p$ now gives the number of O neighbors of A and $q$ gives the number of Zn neighbors of B. Notice that the bond length of any pair reduces with increasing presence of neighboring Zn and O atoms.  Additivity is not applicable for (GaN)(ZnO), not only because of the nonisovalency, but also due to the ionic nature of the constituent semiconductors.  Evidently, there is no natural bond length of GaO and ZnN in the wurtzite structure.  Having no empirical framework to predict the dependence of bond length on local environment, we plot in Fig. 11 the distribution of local bond lengths $d(\rm AB)_{pq}$ as a function of local environment $p,q$.   Notice that the bond length of $\rm {ZnO}_{44}$ and $\rm {GaN}_{00}$ are not too different from their corresponding bond length in bulk semiconductors (indicated by left and right arrows in the figure).  It seems natural to propose that each type of bond has a ``natural'' bond length, which is perturbed both by variation in nearest-neighbor counts and by longer range strains.  For the natural length, we use the average value $<d(\rm AB)_{22}>$ found in the structural relaxation.  Increasing the number of neighbouring Zn and O atoms reduces the length of each bond type by a certain amount $\Delta(AB)_{\rm Zn}$ and $\Delta(AB)_{\rm O}$ separately. Also, notice that $\Delta(AB)_{\rm Zn}$ depends on the neighbouring number of $A$, namely $m$, and vice versa. Thus, the dependence of bond length on local environment is fitted by an empirical model, which neglects the dependence of longer-range strains,
\begin{eqnarray}
\fl d({\rm AB})_{pq}=d({\rm AB})_{22}+(p-2)(q-2)\delta({\rm AB})\nonumber\\
-(q-2)\Delta({\rm AB})_{\rm Zn}-(p-2)\Delta({\rm AB})_{\rm O}
\end{eqnarray}
The fitted parameters $d({\rm AB})_{22}$, $\Delta$ and $\delta$  are listed in Table \uppercase\expandafter{\romannumeral1}.\\
\begin{figure}[htb!]
 \centering
 \includegraphics[scale=0.1]{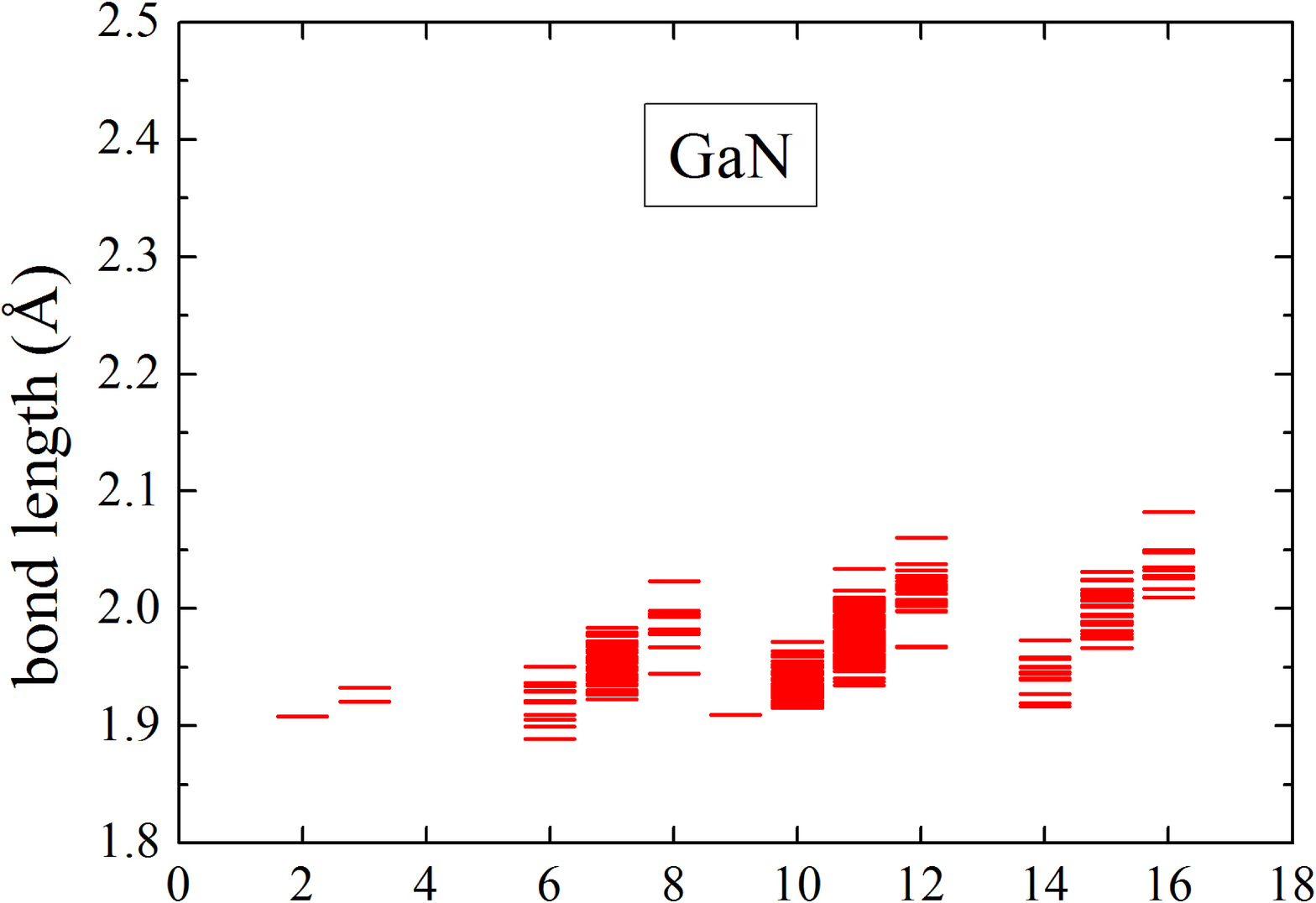}
 \includegraphics[scale=0.1]{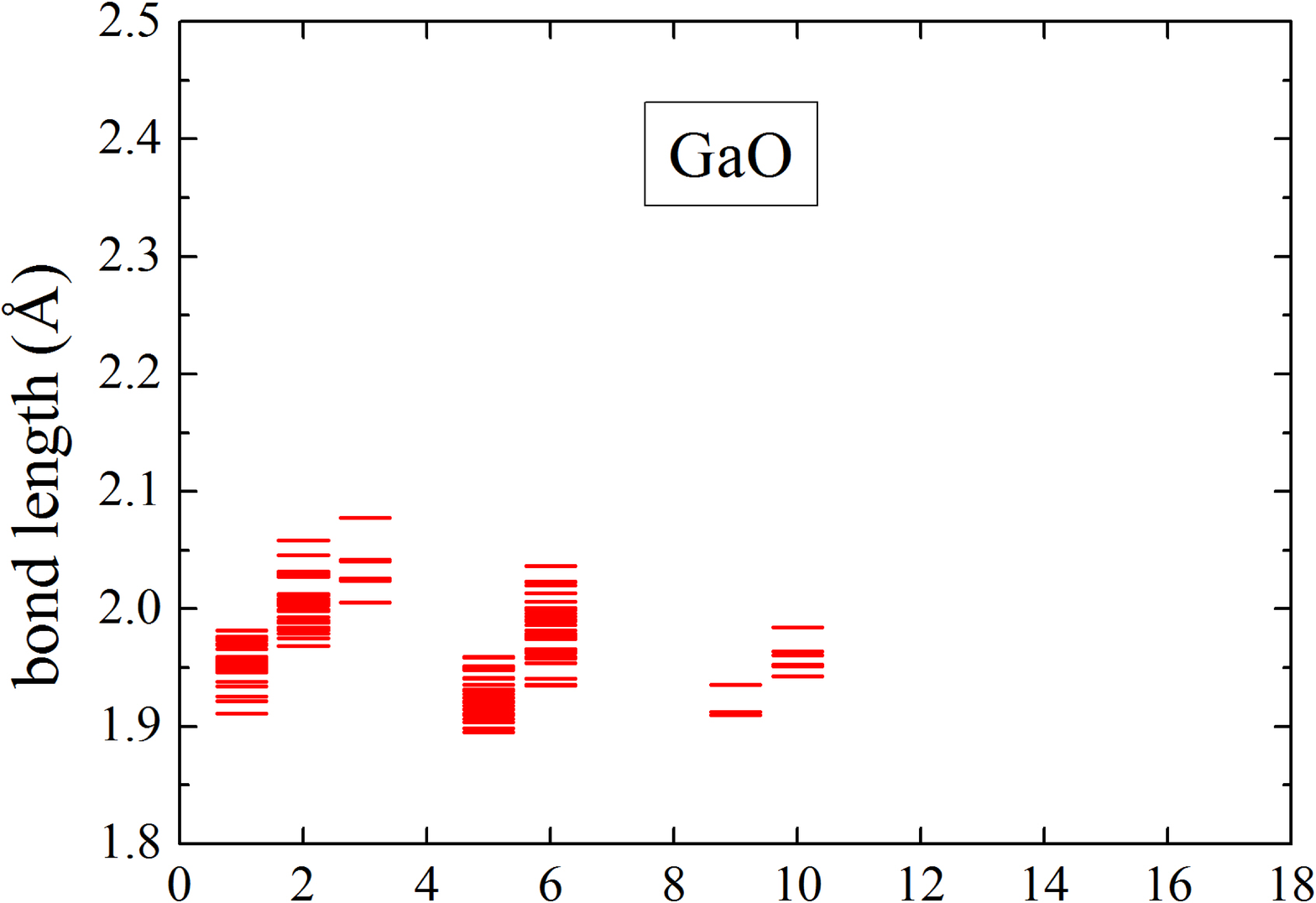}
 \includegraphics[scale=0.1]{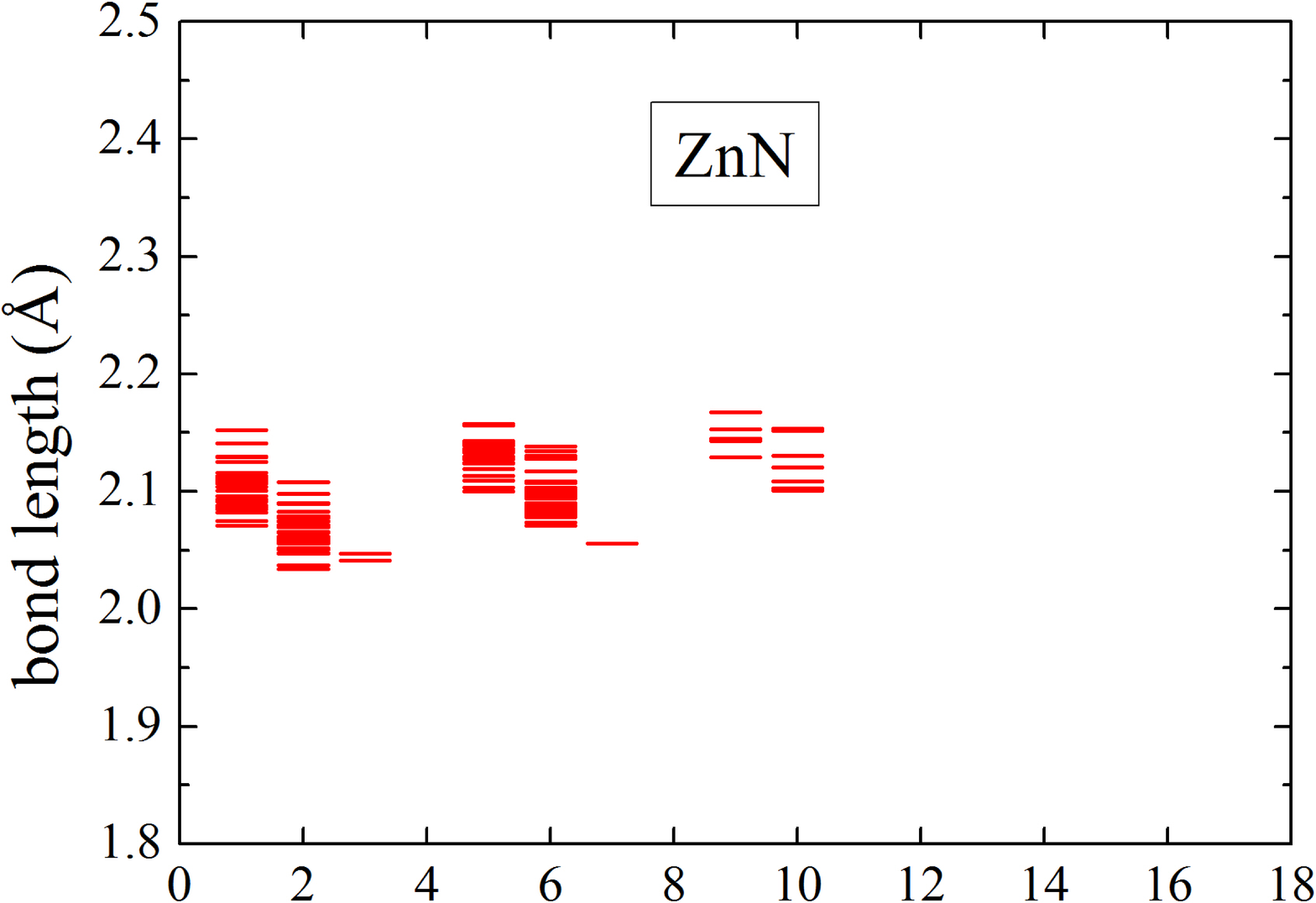}
 \includegraphics[scale=0.1]{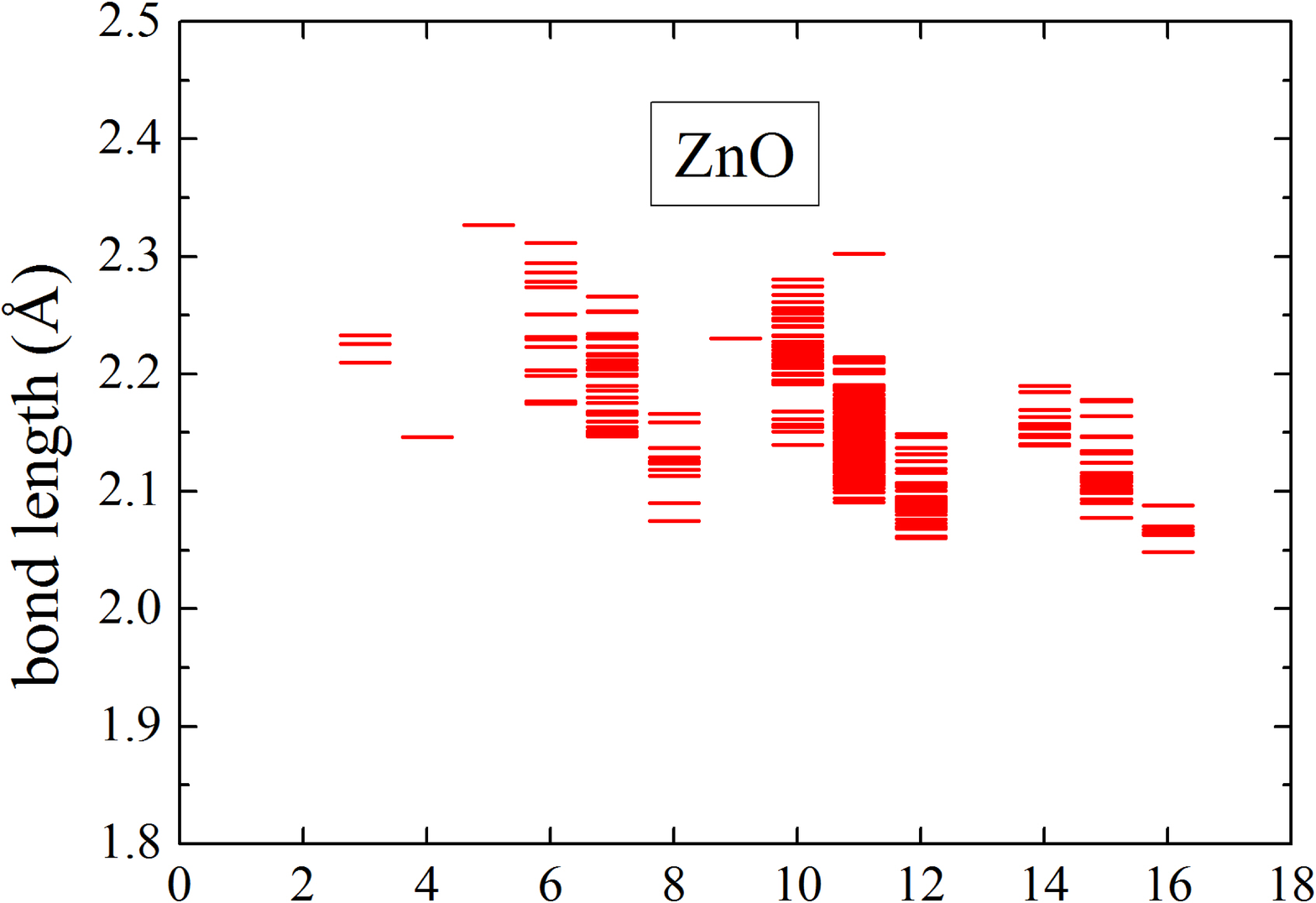}
 \caption{SRO bond-length $d({\rm AB})_{mn}$ for a the $E=0.0295$ eV/atom configuration, after structural relaxation. The $x$-axis, from 1 to 16, gives the neighbor numbers $(mn)$ of the various $AB$ bonds, sequenced as [11-14, 21-24, 31-34, 41-44].  The ZnO bond length is on average larger than other bonds.  Bond length reduces with increasing the number of Zn/O neighbors around anion/cation sites. Notice that Zn and O are the most electropositive and electronegative element in the alloy. The presence of Zn and O in the neighborhood pushes electrons to the anion and pull electrons out of cation separately, increasing the ionicity of the bond, and therefore making the bond length shorter.}
\end{figure}
\begin{figure}[htb!]
 \centering
 \includegraphics[scale=0.2]{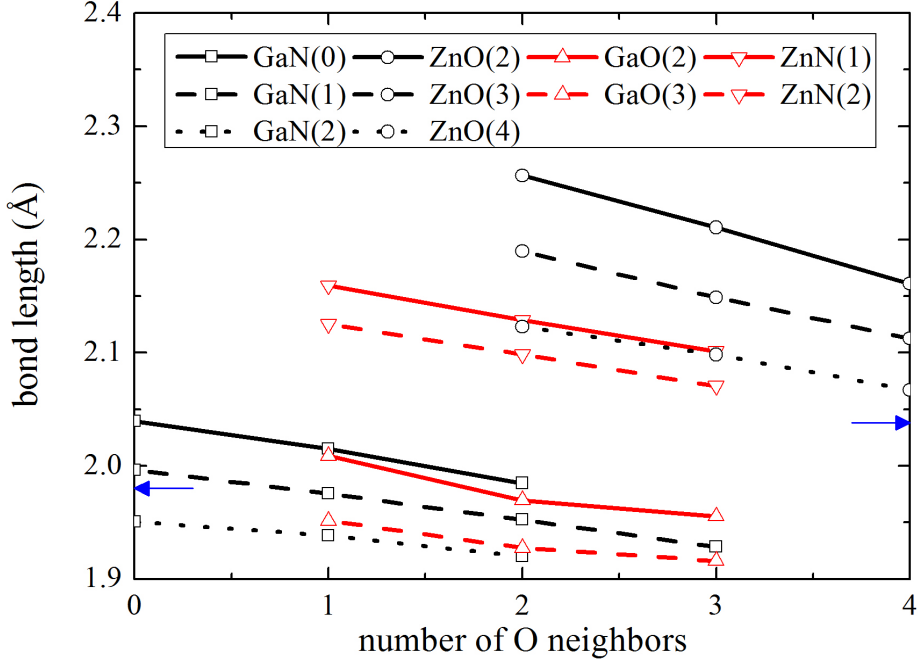}
 \caption{Average cation-anion bond length $\bar{d}({\rm AB})_{pq}$ plotted {\it versus} number of cation O neighbors.  Different numbers of Zn neighbors are plotted as separate lines, as indicated parenthetically in the legend.   The data used are those of Fig. 10.  The bond length reduces with increasing O or Zn neighbors. For example, the ZnO bond length for O having 2, 3 or 4 Zn neighbors  is shown in the upper right by three nearly parallel lines. The arrows indicate the bond lengths of pure GaN ($d=1.98$ $\AA$) and pure ZnO ($d=2.04$ $\AA$) according to DFT.  Statistical errors are the rms deviation taken from Fig. 10, divided by the square root of the number of entries in Fig. 10.  They are mostly too small to plot.}
\end{figure}\\
\begin{table}
\caption{\label{arttype} Empirical model parameters in $\AA$.}
\begin{indented}
\item[]\begin{tabular}{@{}lllll}
\br
&$d(\rm AB)_{22}$&$\Delta_{Zn}$&$\Delta_{O}$&$\delta$\\
\mr
$\rm GaN$&1.9199&0.0324&0.0153&0.0070\\
$\rm GaO$&1.9695&0.0406&0.0268&-\\
$\rm ZnN$&2.0984&0.0316&0.0275&-\\
$\rm ZnO$&2.2565&0.0668&0.0479&0.0095\\
\br
\end{tabular}
\end{indented}
\end{table}

\section{Discussion and Conclusions}

Disordered alloys at low $T$ are non-equilibrium states; the true equilibrium involves phase separation into mixtures of ordered compounds.  Thermodynamics stabilizes disorder (melts the separated compounds) only at temperatures high enough that the entropic term in $F=\bar{E}-TS$ alters the free-energy ($F$) balance.  But alloys result from high $T$ synthesis where ordered phases may not occur.  Then kinetics will prevent equilibration below some temperature $T^\ast$.  In the idealized case of extremely high $T^\ast$, the alloy would be very random, with little short-range order.  The challenge is to find the actual order that freezes in after high $T$ fabrication.  Theory is not very good at this.  We do not have good ``first-principles'' theories of ion diffusivity.  Molecular dynamics, even with approximate empirical potentials, does not permit long-enough simulations to give realistic equilibration.  

The present paper offers a partial way around the difficulties.  We use CE energies carefully fitted to relaxed $T=0$ DFT energies, which are likely to be relatively reliable.  The MC algorithm probably deals accurately with alloy thermodynamics.  Entropy of disorder is correctly included, but vibrational contributions to entropy are omitted.  In many cases, these may not vary too much with alloy order, and can be safely ignored.  
In our case of GaN/ZnO alloys, fabrication is done well above 1000K.  Our MC results show that there is very significant SRO at these temperatures, and that the degree of order varies only slowly with $T$.  Therefore, even though we do not know $T^\ast$, nevertheless, the choice $T^\ast=1200$K probably does not introduce much error.

A definite limitation comes from the finite size of the supercells employed.  The CE parameters at longer range than second neighbors did not improve the fit much\cite{LL}, so supercells with $N\le 192$ are probably sufficient for fitting these parameters.  Our bond-length relaxation study used $N=432$.  One can ask whether the computed bond lengths are affected by the artificial periodic images that the finite supercell imposes.  The relaxations were considered complete when the maximum force on any atom was less than 0.002 eV/\AA.  These forces used DFT kinetic energies and charge densities computed from wave functions at the $\vec{k}=0$ point of the supercell Brillouin zone.  As a test of the influence of periodic replicas, we recomputed, for one of the $x=0.5$ configurations, the interatomic forces with DFT charge densities $\rho$ computed with a $2\times 2\times 2$ Monkhorst-Pack mesh\cite{Monk}.  In other words, the wavefunctions used for $\rho$ were weighted averages over fully periodic states $\psi(\vec{k}=(000))$, and fully anti-periodic states $\psi(\vec{k}=(\frac{1}{2}\frac{1}{2}\frac{1}{2}))$ (a corner of the supercell Brillouin zone), and various intermediate periodicities like $\psi(\vec{k}=(hkl))$ with $h$ and $k$ and $l$ equal to $0$ or $\frac{1}{2}$.  The change of wave-function boundary conditions causes a change of interatomic force, which is a measure of the influence of artificial periodicity.  The maximum force increased to 0.02eV/\AA, with a mean force of 0.005 eV/\AA.  This indicates that the atomic displacements are hardly affected by artificial periodic images. 

In summary, we analyze the strong short-range order in the (GaN)$_{1-x}$(ZnO)$_x$ semiconductor alloy using MC simulation.  The temperature and composition dependence of short-range order and configurational entropy are studied. Short-range order parameters are shown to decay slowly with increasing temperature. Given the short-range order parameters, the statistics of pair distribution is approximated by a binomial distribution.  The configurational entropy is approximated in terms of short-range order parameters.  The dependence of bond-length on local chemical environment is shown to be significant. For any nearest-neighbor cation-anion bond, the bond length reduces with increasing presence of neighbouring Zn and O atoms. Based on the statistics of bond-length for $x$=0.5 at $1200$K, we fit an empirical model which can be used as a starting guess in DFT structural relaxations to reduce the computational effort.\\

\appendix
\section{Configurational entropy and Short-Range Order parameters}

Ref. \cite{Onabe} gives a derivation of the approximate SRO-corrected configurational entropy. Starting from the Warren-Cowley definition (Eq. 1) of the SRO parameter, the probability of finding pairs in the $i^{\rm th}$ neighbor shell, of a simple binary A$_x$B$_{1-x}$ alloy is 
\numparts
\begin{eqnarray}
p_{A \vert B,i}=y(x-x\gamma_i)\\
p_{B \vert A,i}=x(y-y\gamma_i)\\
p_{A \vert A,i}=x(x+y\gamma_i)\\
p_{B \vert B,i}=y(y+x\gamma_i),
\end{eqnarray}
\endnumparts
where $y=1-x$.  We see that $p_{A \vert B,i}=p_{B \vert A,i}$, $p_{B \vert A,i}+p_{A \vert A,i}=x$ and $p_{A \vert B,i}+p_{B \vert B,i}=y$. For an alloy with $N$ atoms, the number of pairs of the $i^{\rm th}$ shell is  $\frac{q_iN}{2}$, with $q_i$ the coordination number of the $i$th shell. The average number of each type is 
\begin{equation}
N_{A \vert B,i}=\frac{q_iN}{2}p_{A \vert B,i},
\end{equation}
and similarly for $N_{B \vert A,i}$, {\it etc.}  The number of ways of arranging this many pairs is in first approximation
\begin{equation}
W^\prime_i(x, N, \gamma_i)=\frac{(\frac{q_iN}{2})!}{N_{A \vert A,i}!N_{A \vert B,i}!N_{B \vert A,i}!N_{B \vert B,i}!}
\end{equation}
However, $W^\prime_i(x, N, \gamma_i)$ contains unphysical arrangements, where one site is occupied by both $A$ and $B$ atoms. Therefore, the exact enumeration $W_i(x, N, \gamma_i)$ is always smaller than $W^\prime_i(x, N, \gamma_i)$. We assume a universal function $h(x,N)<1$ defined by $W_i(x, N, \gamma_i)=h(x,N)W^\prime_i(x, N, \gamma_i)$. We choose to fix this ``universal function'' by the relation $h(x,N)=W_i(x, N, \gamma_i^0)/W^\prime_i(x, N, \gamma_i^0)$, where $\gamma_i^0$ is the value of $\gamma_i$ which maximizes $W_i(x, N, \gamma_i)$.  Further, we  approximate the maximum by an upper limit, its summation, 
\begin{equation}
W_i(x, N, \gamma_i^0)\approx\sum_{\gamma_i}{W_i(x, N, \gamma_i)}.
\end{equation}
The summation on right-hand side is just the total number of possible alloy configurations, $W(x, N)=\frac{N!}{N_A!N_B!}$.  Similarly, we approximate the numerator $W_i'(x, N, \gamma_i^0)$  of $h(x,N)$ by an upper limit, its enumeration in the completely random alloy, namely
\begin{equation}
W_i'(x, N, \gamma_i^0)=\frac{(\frac{q_iN}{2})!}{N_{A \vert A,i}^0!N_{A \vert B,i}^0!N_{B \vert A,i}^0!N_{B \vert B,i}^0!},
\end{equation}
where $N_{A \vert B,i}^0=\frac{q_i}{2}\frac{N_AN_B}{N}$, etc. Therefore the universal function $h(x,N)$ is approximated by
\begin{equation}
h(x,N)=\frac{N!}{N_A!N_B!}\frac{N_{A \vert A,i}^0!N_{A \vert B,i}^0!N_{B \vert A,i}^0!N_{B \vert B,i}^0!}{(\frac{q_iN}{2})!}.
\end{equation}
The entropy then can be approximated in the thermodynamic limit using Stirling's approximation
\begin{eqnarray}
\fl k_B\ln W_i(x, N, \gamma_i)=k_B\ln h(x,N)+k_B\ln W_i'(x, N, \gamma_i)\nonumber\\
=N\{S_{rand}-q_iS_{rand}-k_B\frac{q_i}{2}[x(x+y\gamma_i)\ln x(x+y\gamma_i)+\nonumber\\
y(y+x\gamma_i)\ln y(y+x\gamma_i)+2xy(1-\gamma_i)\ln xy(1-\gamma_i)]\}.
\end{eqnarray}

\ack
We are grateful for discussions with members of the SWaSSiT group including Neerav Kharche, James T. Muckerman and Mark S. Hybertsen. This research used computational resources at the Center for Functional Nanomaterials, Brookhaven National Laboratory, which is supported by the US Department of Energy under Contract No. DE-AC02-98CH10886. Work at Stony Brook was supported by US DOE Grant No. DE-FG02-08ER46550 (PBA) and DE-FG02-09ER16052 (MFS).

\end{document}